\documentclass[twocolumn,sort&compress,noshowpacs,preprintnumbers,amsmath,amssymb,aps,floatfix,prl]{revtex4}
 
\usepackage{graphicx}
\usepackage{epstopdf}
\usepackage{amsmath}
\usepackage{multirow}
\usepackage{times}
\usepackage{amssymb}
\usepackage{dcolumn}%
\usepackage{bm}%
\usepackage{upgreek}
\usepackage{wasysym}
\usepackage{verbatim}
\usepackage{blindtext}
\usepackage[driverfallback=dvipdfm]{hyperref}

\hypersetup
{
	colorlinks=true,
	linkcolor=blue,
	citecolor=blue
}

\begin{document}

\title{Molecular layers in thin supported films exhibit the same scaling as the bulk between slow relaxation and vibrational dynamics}

\author{M.\@ Becchi}
\affiliation{Dipartimento di Fisica ``Enrico Fermi'', 
Universit\`a di Pisa, Largo B.\@Pontecorvo 3, I-56127 Pisa, Italy}

\author{A.\@ Giuntoli}
\affiliation{Dipartimento di Fisica ``Enrico Fermi'', 
Universit\`a di Pisa, Largo B.\@Pontecorvo 3, I-56127 Pisa, Italy}

\author{D.\@ Leporini}
\email{dino.leporini@unipi.it}
\affiliation{Dipartimento di Fisica ``Enrico Fermi'', 
Universit\`a di Pisa, Largo B.\@Pontecorvo 3, I-56127 Pisa, Italy}
\affiliation{IPCF-CNR, UOS Pisa, Italy}

\begin{abstract}
\noindent
We perform molecular-dynamics simulations of a supported molecular thin film. 
By varying thickness and temperature, we observe anisotropic mobility as well as strong gradients of both the vibrational
motion and the structural relaxation through film layers with monomer-size thickness. 
We show that the gradients of the fast and the slow dynamics across the layers (except the adherent layer to the substrate)
comply, without any adjustment, with the same scaling between the structural relaxation time and the
Debye–Waller factor originally observed in the bulk [Larini et al., Nat. Phys., 2008, 4, 42]. 
The scaling is not observed if the average dynamics of the film is inspected. Our results suggest that the solidification
process of each layer may be tracked by knowing solely the vibrational properties of the layer and the bulk.
\end{abstract}


\maketitle

\section{Introduction}
Ultra-thin organic soft films made by small molecular weight or polymeric units are involved in tissue engineering \cite{LiuTissueEngin2004}, mechanically flexible electronics \cite{PeumansSmallMWOrganThinFilmJApplPhys03},  memories \cite{YangMemoriesAFM} and displays  \cite{HowardDisplayOrganicFilmSciAmer04}. Mechanical flexibility, scalability to the nanoscale and processability are some of their most appealing features. 
In supported thin films the mobility is slowed down close to the  solid substrate \cite{GranickLangm96,NapolitNatCom11,BinderPaulThreeStepDecayEPL12} and enhanced at the free interface 
\cite{YangSurfaceLayerScience10,ForrestACSMacro14} with considerable distribution of the solidification temperature, i.e. the glass transition (GT) temperature $T_g$ \cite{TorkelsonNMat03}.
The differences between soft thin films and their bulk counterpart are widely reported by dedicated topical reviews \cite{RothFilm,ForrestFilm,sim,JonesFilm,McKennaFilm,NapReview}. 

Molecular reorganization deals with long-time transport properties. Yet, several experimental and numerical studies in {\it bulk} evidenced universal correlations between the long-time structural relaxation and the fast (picosecond) dynamics as sensed by the Debye-Waller (DW) factor $\langle u^2\rangle$, the rattling amplitude of the particle within the cage of the first neighbours  \cite{HallWoly87,OurNatPhys,lepoJCP09,Puosi11,UnivPhilMag11,OttochianLepoJNCS11,UnivSoftMatter11,Puosi12SE,DouglasCiceroneSoftMatter12,SpecialIssueJCP13,CommentSoftMatter13,SokolovNovikovPRL13, TVG1,Merabia_JCP17,Vogel_JCP17,Puosi18SE}.
 In particular, correlations are found in polymers \cite{OurNatPhys,lepoJCP09,Puosi11,Puosi18SE}, binary atomic mixtures \cite{lepoJCP09,SpecialIssueJCP13}, colloidal gels \cite{UnivSoftMatter11}, antiplasticized polymers \cite{DouglasCiceroneSoftMatter12}, water \cite{SokolovNovikovPRL13} and water-like models \cite{Merabia_JCP17,Vogel_JCP17}.
 
 {The extension of the correlation between the vibrational dynamics and the slow relaxation observed in bulk to  thin films \cite{DouglasStarrPNAS2015} - also in connection to the strictly related theme of the changes of $T_g$\cite{SimmonsConfined_u2_ACSMacroLett14} - has been investigated by numerical studies considering the {\it average} dynamics of the film.}  
Nonetheless, owing to the strong gradients of mobility and relaxation \cite{TorkelsonNMat03}, high-resolution studies are needed and it has been noted \cite{Hanakata2012Film} that, since the spatial variation of relaxation is difficult to access experimentally, the Debye-Waller factor may be an effective measure for probing spatial variations of relaxation through the film.
 
Motivated by the previous remarks we carried out a thorough campaign of molecular dynamics (MD) simulations of a  supported molecular thin film.  Here, we show that, { with the exception of the layer of { monomers} adhering to the substrate \cite{GranickLangm96,NapolitNatCom11,BinderPaulThreeStepDecayEPL12}}, 
 the gradients of the DW factor $\langle u^2\rangle$ and the structural relaxation time across film layers  with {  \it monomer-size} thickness correlate strongly and exhibit the {\it same}  master curve observed in {\it bulk} with {\it no} adjustable parameter \cite{OurNatPhys}.

\section{Simulation Methods}
\label{SimMet}
We model molecules as unentangled linear chains of beads linked
by harmonic springs. The substrate supporting the polymer film is modeled as a
collection of substrate atoms and coupled to the chains. The film is under vacuum, i.e. no pressure is exerted. { Each linear chain has $M=3$ monomers}. Non-bonded monomers belonging to the same or different chains interact with a truncated Lennard-Jones (LJ) potential:
\begin{equation}
\label{eq1}
U^{LJ}(r)=\varepsilon\left [ \left (\frac{\sigma^*}{r}\right)^{12 } - 2\left (\frac{\sigma^*}{r}\right)^6 \right]+U_{cut}
\end{equation}
$\sigma^*=2^{1/6}\sigma$ is the position of the potential minimum with depth $\varepsilon$. The value of the constant $U_{cut}$ is chosen to ensure $U^{LJ}(r)=0$ at $r \geq r_c=2.5\,\sigma$. Bonded monomers interact with an harmonic potential $U^b(r)=k(r-r_0)^2$ with $k=555.5 \, \varepsilon  / \sigma^2$ and $r_0=0.9\,\sigma$. Henceforth, all quantities are expressed in terms of reduced Lennard-Jones units, i.e. $\varepsilon = 1$, $\sigma=1$,  with unit monomer mass and Boltzmann constant. The reduced units can be
mapped onto physical units relevant to generic non-equilibrium fluids, by taking MD time, length and energy units as corresponding roughly to about $2$ ps, $0.5$ nanometer and $3.7$ kJ/mol , respectively \cite{Kroger04}. 

To model the substrate, we tether each substrate atom to one site of a square lattice at $z=0$ spaced by $0.9 \cdot 2^{1/6}$ with harmonic potential $U^s(r)=k_s r_s^2$ where $k_s=100$ and $r_s$ denotes the distance between the substrate atom and the tied site of the lattice. Substrate atoms are not mutually interacting, whereas they are coupled to the polymer monomers with the same LJ potential of the non-bonded monomers. Molecular-dynamics (MD) numerical simulations were carried out with the LAMMPS code (http://lammps.sandia.gov) \cite{PlimptonLAMMPS}. 
The two transversal dimensions of the film are considered as infinite and periodic boundary conditions are applied. We simulated samples with different temperatures (0.47, 0.48, 0.49, 0.5)  and number of total monomers N = 2001, 3000, 3999 corresponding to different film thicknesses of about  5, 7.5 e 10 according to the criterion of ref. \cite{Hanakata2015Film}.
Additionally, we simulate a bulk system with M = 3 and N = 3999 at zero pressure for the purpose of comparison. All the systems where initially equilibrated in the $NPT$ ensemble (constant number of particles, pressure and temperature) with $P=0$ to allow full correlation loss of the end-end vector of the polymer chains ($\lesssim 0.1$). Production runs were carried out in the $NVE$ ensemble (constant number of particles, volume and energy). Up to sixty-four independent replicas of each state were considered to ensure suitable statistical average.

\section{Results and Discussion}
We present and discuss results concerning the simulation of the microscopic dynamics of a thin molecular film supported by a solid substrate.

\subsection{Spatial variation of the density}
\begin{figure}[t]
\centering
\includegraphics[width= 0.9 \linewidth]{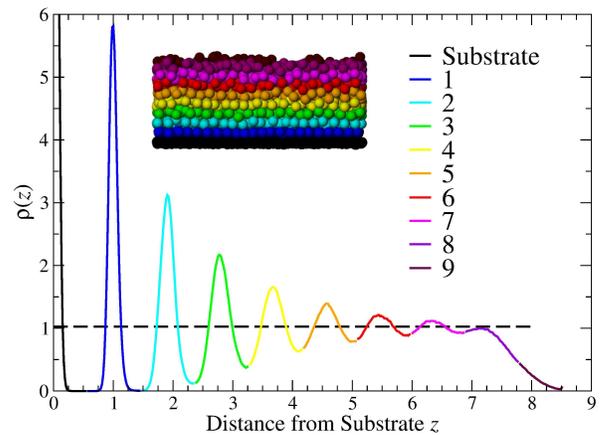}
\caption{Density profile of the thin molecular film with thickness $h=7.5$ at $T=0.49$. The film is supported by a substrate located at $z=0$ (continuous black line). Monomers are organized in well-defined layers of thickness about $0.9\sigma$, i.e. about the monomer size. The layers are labelled by a suitable colour code and their exact definition is given in the text. The dashed line ($\rho=1.024$) represents the density of the equivalent bulk system at the same temperature and pressure ($P=0$). {The picture shows a configuration of the film.} }
\label{fig1}
\end{figure}

Fig.\ref{fig1} shows the density of the molecular film with thickness $h=7.5$ and temperature $T=0.49$  at a distance $z$ from the substrate (thickness measured according to ref. \cite{Hanakata2015Film}). It is apparent that the presence of the latter favours the organisation of the particles in well-defined layers with width comparable to the monomer size, about $\sigma = 1$. This is a marked difference with respect to the corresponding bulk system where density is homogeneous. To analyse the behaviour of each layer, we partition the particles of the film in subsets. Each subset corresponds to a  single layer with width $0.9$. More precisely, having located the density maximum of the substrate at $z=0$, the i-th particle with distance $z_i $ from the substrate  belongs to the m-th layer if  {$0.9(m-1/2)+0.05 \leq z_i \leq 0.9(m+1/2)+0.05$}. 
The result of the partition is shown in Fig.\ref{fig1} in terms of a suitable color code.

\subsection{Spatial variation of mobility and relaxation}
The spatial distribution of the particle packing across the film leads to a corresponding spatial distribution of both the mobility and the relaxation. The matter is discussed in this Section.

\subsubsection{Mobility}

First, we investigate the mobility of the particles {\it initially} located in the m-th layer. To this aim, we define their mean square displacement (MSDm):
\begin{equation}
<r^2(t)>_m = \Big \langle \frac{1}{N_m}   \sum_{j=1}^{N_m} [{\bf r}_j(t) - {\bf r}_j(0)]^2  \Big \rangle
\label{MSDm}
\end{equation}
${\bf r}_j(t)$ is the position of the j-th particle at time $t$. The latter was at the initial time one of the $N_m$ particles of the m-th layer.  $\langle \dots \rangle$ denotes the average over the system replicas to improve the statistical precision. 

\begin{figure}[t]
\centering
\includegraphics[width=\linewidth]{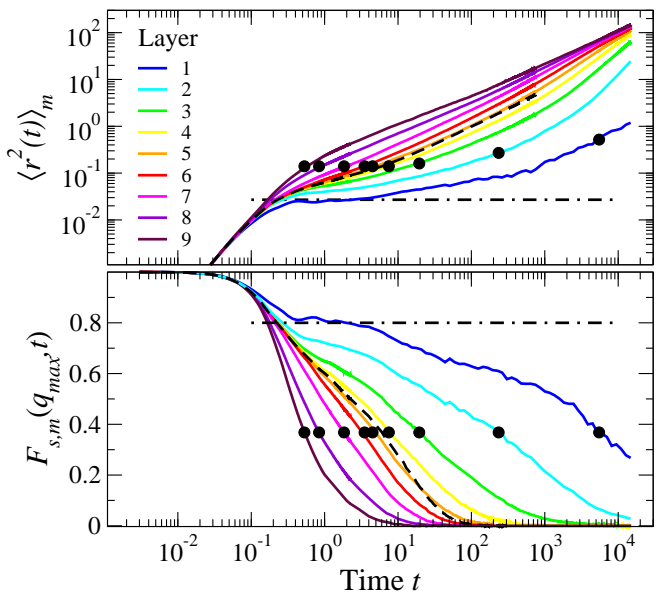}
\caption{ MSDm, Eq.\ref{MSDm}, (top) and ISFm , Eq.\ref{ISFm}, (bottom) of the particles initially belonging to the layers of the film with thickness $h=7.5$ at $T=0.49$ shown in Fig.\ref{fig1} (same color code). The black dot-dashed lines show the plateau levels reached by the two quantities for particles belonging to the substrate. It is seen that the layer adherent to the substrate (blue line)  is strongly coupled to the substrate up to time $t \sim 4$. Black dots mark the structural relaxation time  of the m-th layer, $\tau_{\alpha, m}$. The  black dashed lines are the MSD and ISF curves of the bulk system at the same temperature and pressure of the film, $T=0.49$ and $P=0$, respectively.
}
\label{fig2} 
\end{figure}

Representative results concerning the MSDm of all the layers of the film with density profile plotted in Fig.\ref{fig1} are shown in Fig.\ref{fig2} (top panel). To appreciate the huge spread due to the film confinement, the mobility of the equivalent bulk system is superimposed. Going into details, one sees that for very short times, $t \lesssim 0.1$, the MSDm increase is independent of the layer since the particle displacement is ballistic, i.e. $<r^2(t)>_m \simeq 3 T t^2$. { 
Later, MSDm increases less due to two distinct effects: i) the trapping due to the cage of the surrounding particles, and ii) the absorption-desorption process of the particles close to the substrate. While the cage effect slows down all the particles, the absorption-desorption process is felt, via the molecular connectivity, only at short distance from the substrate where it lowers the mobility. We remind that the substrate is solid, i.e. its particles perform small-amplitude random oscillations around their average positions so that after the ballistic regime their MSD reaches a plateau with no further increase. Fig.\ref{fig2} (top panel) shows a {\it multiple} transient arrest in the closest layer to the substrate which, for conciseness reasons, will be denoted henceforth as the "adherent layer". 
The multiple transient arrest of the particles of the adherent layer is evidenced by a first plateau with MSD coinciding with the substrate one in the time window $\sim 0.4-4$, followed by a mild increase, and a later plateau in the time window $\sim 20 -90$.  That scenario have been already observed at polymer-solid interfaces with attractive interactions \cite{BinderPaulThreeStepDecayEPL12,ZhangDouglasStarrThinSupportedFilmPNAS18}. The first plateau signals the arrest due to the adsorption, whereas the second one pertains to desorbed particles awaiting for the escape from the cage of the neighbours. At later times, the escape is seen in all the layers. Early escape events yield a change of the concavity of the time dependence of MSDm and  the presence of an inflection point at $t^*$. By performing the same analysis as in ref. \cite{OurNatPhys} one finds $t^* \simeq 1$ as in  bulk systems.}  MSDm at $t^{\star}$ is interpreted as a mean localization length and we define the DW factor of the m-th layer as $\langle u^2 \rangle_m \equiv  \langle r^2(t=t^{\star})\rangle_m$. 
For more details, see Ref. \cite{OurNatPhys}. For times fairly longer than $t^*$ the mobilities of the particles belonging {\it initially} to different layers tend to equalise. This is due to the fact that at long times particles move from one layer to the other ones leading to an averaged mobility across the film.  

\subsubsection{Structural relaxation}

The structural relaxation following the escape process of one particle {\it initially} belonging to the m-th layer from the cage of the first neighbouring particles is conveniently described by the self part of the intermediate scattering function (ISFm) \cite{HansenMcDonaldIIIEd}:
\begin{equation}
F_{s,m}(q, t) =  \Big \langle \frac{1}{N_m}   \sum_{j=1}^{N_m}  exp \{-i {\bf q}\cdot [{\bf r}_j(t) - {\bf r}_j(0)] \} \Big \rangle
\label{ISFm}
\end{equation}
ISFm is evaluated at the wavevector $q=q_{max}$ with $q_{max}$ being the q-vector of the maximum of the static structure factor corresponding to about the distance of nearest-neighbours. By construction, ISFm is negligibly small when the particle displacement exceeds a few particle diameters. Fig.\ref{fig2} (bottom panel) plots ISFm of the film with density profile plotted in Fig.\ref{fig1}.  It is seen that the ISFm decay is identical in the different layers in the short-time ballistic regime. Later, the decay is slowed down in a way depending on the layer and paralleling the progress of mobility shown in Fig.\ref{fig2} (top panel). In particular, the adherent layer exhibits a temporary arrest of the relaxation at the same height of the substrate in the time window $\sim 0.4-4$ followed by a second plateau in the time window $\sim 20 -90$ and a further decay in agreement with previous studies \cite{BinderPaulThreeStepDecayEPL12,ZhangDouglasStarrThinSupportedFilmPNAS18}.

\subsubsection{The layer structural relaxation time $\tau_{\alpha, m}$}

\begin{figure}[t]
\centering
\includegraphics[width=0.9 \linewidth]{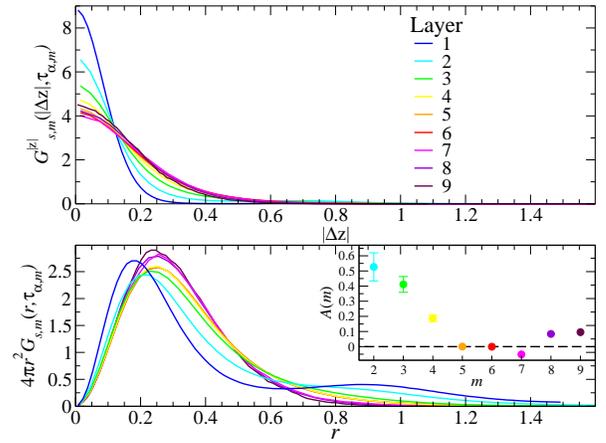}
\caption{Displacement distributions in a time $\tau_{\alpha, m}$ of a particle initially in the m-th layer of the film with thickness $h=7.5$ at $T=0.49$ shown in Fig.\ref{fig1} (same color code). Top: distribution of the absolute value of the changes of elevation from the substrate, Eq.\ref{VHmodz}. The half width of a layer is about $0.45$. Bottom: van Hove distribution of the distance between the initial and the final position, Eq.\ref{VH}.  Inset: anisotropy function $A(m)$, Eq.\ref{Am}. Layers close to the substrate and the free interface have larger intra-layer  than inter-layer mobility. {The anisotropy of the adherent layer is very large, $A(1) \gg 1$, and not shown for clarity reasons. Error bars are indicated only if larger than the dot size.}
}
\label{fig3} 
\end{figure}

We are interested in the definition of the characteristic structural relaxation time of the m-th layer, $\tau_{\alpha, m}$.  A convenient definition has to ensure that within $\tau_{\alpha, m}$ particle rearrangements relax effectively the cage structure with displacements along the z direction not exceeding the half-layer thickness. Fulfilling these requirements  is not obvious since the molecular film is partitioned in layers as thin as about one particle diameter. Nonetheless, we remind the well- known result that the atomic MSD during the structural relaxation is less than one atomic radius \cite{Angell91}.  We find that a proper definition of $\tau_{\alpha, m}$ is the familiar one defined by the equation $F_{s,m}(q_{max},\tau_{\alpha, m})=1/e$. 
{To motivate this choice, we define the distribution function:
\begin{equation}
G_{s,m}^{z}(\Delta z, t)=\Big \langle \frac{1}{N_m}  \sum_{j=1}^{N_m} \delta\left[\Delta z -z_j(t)+z_j(0)\right] \Big \rangle
\label{VHz}
\end{equation}
where $\delta[\cdots]$ and $z_j(t)$ are the Dirac delta and the elevation of the j-th particle from the substrate at time $t$, respectively. At the initial time the j-th particle is one of the $N_m$ particles of the m-th layer. The quantity $G_{s,m}^{z}(\Delta z, t) d \Delta z$ is the  probability that the particle initially in the m-th layer changes the initial distance from the substrate between $\Delta z$ and  $\Delta z+d\Delta z$ after a time $t$.  We are interested in the distribution of the modulus of $\Delta z$, 
\begin{equation}
G_{s,m}^{| z |}(|\Delta z |, t) = G_{s,m}^{z}( |\Delta z |, t) + G_{s,m}^{z}(-| \Delta z |, t)
\label{VHmodz}
\end{equation}
which is normalised in the positive semiaxes $ |\Delta z | \ge 0$.

The top panel of Fig.\ref{fig3} shows the distribution $G_{s,m}^{| z |}(|\Delta z |, \tau_{\alpha, m})$ of all the layers of the film with thickness $h=7.5$ at $T=0.49$. It is seen that the elevation change in a time $\tau_{\alpha, m}$ is comparable to or less than the particle radius, i.e. the half-layer thickness.  Identical conclusions are reached by considering the layers of all the films at the different temperatures examined in the present paper. This provides evidence that $\tau_{\alpha, m}$ is a characteristic relaxation time of the m-th layer. Alternative choices for $\tau_{\alpha, m}$ leading to longer time scales are anticipated to be affected by particle exchange between nearby layers so that we think that $\tau_{\alpha, m}$ is a convenient definition of the relaxation time of the m-th layer.

\subsubsection{Anisotropy of the particle displacement in a time  $\tau_{\alpha, m}$}

The  particle displacement in the film is anisotropic  in a time  $\tau_{\alpha, m}$. To show that, we
consider the self part of the van Hove function {\it restricted} to particles which belong to the m-th layer at the initial time:
\begin{equation}
G_{s,m}({\bf r}, t)=\Big \langle \frac{1}{N_m}  \sum_{j=1}^{N_m} \delta\left[{\bf r} -{\bf r}_j(t)+{\bf r}_j(0)\right] \Big \rangle
\label{VH}
\end{equation}
where ${\bf r}_j(t)$ is the position of the j-th particle at time $t$, respectively. At the initial time the j-th particle is one of the $N_m$ particles of the m-th layer.  { We average the distribution $G_{s,m}({\bf r}, t)$ over a spherical shell of radius $r$ and thickness $dr$ to get the spherical van Hove distribution $G_{s,m}(r,t)$.}
The interpretation of $G_{s,m}(r,t)$ is 
direct. The product $G_{s,m}(r,t) \cdot 4\pi r^{2} dr$ is the probability that the particle, initially in the m-th layer, is at a distance between $r$ and $r+dr$ from  the initial position after a time t.}

The bottom panel of Fig.\ref{fig3} shows the van Hove distribution $G_{s,m}(r,\tau_{\alpha, m})$ of a particle initially located in the m-th layer of the  film with thickness $h=7.5$ at $T=0.49$. By comparison with the top panel of  Fig.\ref{fig3}, it is seen that in a time $\tau_{\alpha, m}$ the distance between the initial and the final positions is larger than the change of distance from the substrate which is of the order of the particle radius, i.e. the half-layer thickness. In particular, the self part of the van Hove function shows that the particles  closer to the substrate undergo solid-like large jumps of the order of the particle size, well seen in the peak at $r \sim 1$ which is, instead, virtually missing in the elevation distribution distribution $G_{s,m}^{| z |}(|\Delta z |, \tau_{\alpha, m})$, Fig.\ref{fig3} (top panel). 
That findings suggest that the particle motion is {\it anisotropic} with larger intra-layer displacements with respect to the interlayer ones. The feature is also apparent in the MSD evaluated at 
$\tau_{\alpha, m}$, see  Fig.\ref{fig2} (top panel), which {\it increases} for layers closer to the substrate whereas the corresponding interlayer displacements {\it decrease}, see Fig.\ref{fig3} (top panel). This suggests that, at least in some layers, the structural relaxation of the layer in a time $\tau_{\alpha, m}$ is facilitated by quasi-bidimensional intra-layer displacements. 

To better scrutinize the anisotropy motion we define the anisotropy function $A(m)$ of the m-th layer as:
\begin{equation}
A(m)= \frac{ \langle r^2(\tau_{\alpha, m})\rangle_m}{3 \langle (\Delta z)^2(\tau_{\alpha, m})\rangle_m } - 1
\label{Am}
\end{equation}
where $\langle (\Delta z)^2(\tau_{\alpha, m})\rangle_m$ is the second moment of the distribution of Eq.\ref{VHz} at $t=\tau_{\alpha, m}$. The function $A(m)$ is small if the displacement is isotropic and positive if the intra-layer mobility exceeds the inter-layer mobility. The inset in the bottom panel of Fig.\ref{fig3} shows that the anisotropy is meaningful close to the substrate, { absent in the central region of the film  and weak close to the free surface. The finding is ascribed to the presence of increasing dimensional constraints far from the inner part of the film. We anticipate stronger bi-dimensional character of the motion close to the solid substrate  than to the free interface which has more diffuse character along the $z$ direction, see Fig.\ref{fig1}}.

\subsection{Scaling between slow relaxation and vibrational dynamics}

In bulk systems pioneering studies \cite{HallWoly87} and later investigations involving MD simulations and extended comparison with the experiment \cite{OurNatPhys,SpecialIssueJCP13,TVG1} revealed the strong correlation between the fast vibrational dynamics, characterized by the DW factor $\langle u^2 \rangle$, and the structural relaxation time. {The dynamics was varied by changing several parameters like, e.g., temperature, pressure, inter- and intra-molecular potential and polymer size in one- or two- components systems}. { Notice that, here, the dynamics is changed by resorting to {\it completely different} variables, e.g. the film thickness and the position of the layer, in addition to the temperature.}
The correlation between structural relaxation and fast mobility is summarized by  the master curve \cite{OurNatPhys}:
\begin{eqnarray}
\label{eqn:u2tauExp0}
\log \tau_\alpha &=& \mathcal{F}_{FM}(\langle u^2\rangle) \\ 
\label{eqn:u2tauExp}
&=& \alpha+\tilde{\beta} \, \frac{\langle u_g^2\rangle}{\langle u^2\rangle} +\tilde{\gamma}\left (\frac{\langle u_g^2\rangle}{\langle u^2\rangle} \right)^2
\label{eqn:u2tauExp2}
\end{eqnarray}
$ \langle u^2_g\rangle$ is the fast mobility at GT,
$\tilde{\beta}$ and $\tilde{\gamma}$ are suitable universal constants independent of the kinetic fragility \cite{OurNatPhys,SpecialIssueJCP13}, and $\alpha =  2 - \tilde{\beta} - \tilde{\gamma}$ to comply with the usual definition $\tau_\alpha = 100$ s at the glass transition. For the present molecular model in {\it bulk} systems Eq.\ref{eqn:u2tauExp} reduces to \cite{OurNatPhys}:
\begin{equation}
\log \tau_\alpha = {\alpha} + {\beta} \, \frac{1}{\langle u^2\rangle} + {\gamma} \, \frac{1}{\langle u^2\rangle^{2}}
\label{parabolaMD}
\end{equation}
with $\alpha =-0.424(1), \beta = 2.7(1) \cdot 10^{-2}, \gamma =  3.41(3) \cdot 10^{-3}$. 

\begin{figure}[t]
\centering
\includegraphics[width=\linewidth]{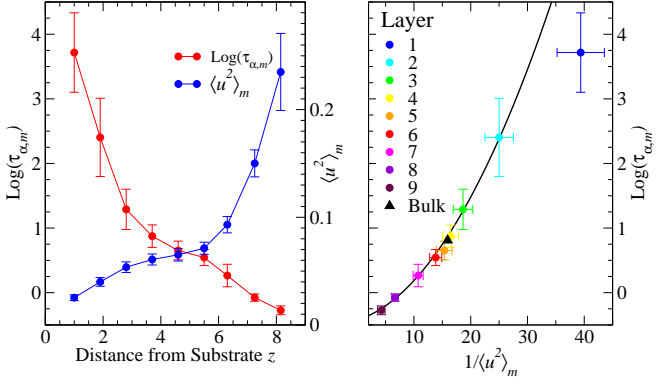}
\caption{{Left: Structural relaxation time and DW factor as a function of the distance $z$ from the substrate of the film with thickness $h=7.5$ at $T=0.49$ shown in Fig.\ref{fig1}. Three regions are seen: a less mobile region close to the substrate - the so called "bound layer" \cite{ZhangDouglasStarrThinSupportedFilmJCP17,ZhangDouglasStarrThinSupportedFilmPNAS18} - , an intermediate bulk-like region, and a more mobile region close to the free interface. Right: correlation between the structural relaxation time and the inverse of the DW factor of the different layers of the film (same color code as in Fig.\ref{fig1}). The black triangle is the state point of the corresponding bulk state with $T=0.49$, $P=0$. The superimposed black curve is Eq.\ref{parabolaMDfilm}. Notice that the adherent layer deviates from the master curve.}}
\label{fig4}
\end{figure}

Douglas and coworkers  developed a localization model predicting the alternative master curve $\mathcal{F}_{FM}(\langle u^2\rangle) \propto \langle u^2\rangle^{-3/2}$ relating the structural relaxation time and the fast mobility  \cite{DouglasCiceroneSoftMatter12,DouglasStarrPNAS2015}. Both the latter form and Eq.\ref{eqn:u2tauExp} account for the convexity of the master curve, evidenced by experiments and simulations, and improve the relation originally proposed  by Hall and Wolynes \cite{HallWoly87}. 

Our claim in the present work is that the scaling form given by Eq.\ref{parabolaMD}, originally found in bulk systems, also works as a master curve of the relaxation time and DW factor of the m-th layer, i.e. we anticipate
\begin{equation}
\log \tau_{\alpha, m} = {\alpha} + {\beta} \, \frac{1}{\langle u^2\rangle_m} + {\gamma} \, \frac{1}{\langle u^2\rangle_m^{2}}
\label{parabolaMDfilm}
\end{equation}
where $\alpha, \beta$ and $\gamma$ are the {\it same} of bulk systems. To start with, the left panel of Fig. \ref{fig4} shows the distribution of both the relaxation time $\tau_{\alpha, m}$ and the DW factor across the molecular film with thickness $h=7.5$ at $T=0.49$. The relaxation is faster, and the DW is larger, on approaching the film interface at large $z$ values. The distribution of the relaxation times extends over about four orders of magnitude. Three regions are seen: a less mobile region close to the substrate - the so called "bound layer" \cite{ZhangDouglasStarrThinSupportedFilmJCP17,ZhangDouglasStarrThinSupportedFilmPNAS18} - , an intermediate bulk-like region, and a more mobile region close to the free interface.

The right panel of Fig. \ref{fig4} is a correlation plot between the relaxation time and the inverse DW factor. The superimposed curve is Eq. \ref{parabolaMDfilm}, i.e. the master curve of bulk systems. As a consistency check, we show that the pair ($\tau_{\alpha}$, $1/\langle u^2\rangle$) of the corresponding bulk molecular liquid at same pressure $P=0$ and temperature fulfills the scaling. It is seen that the corresponding pairs of all the layers, but the single adherent layer, do the same. The scaling holds even for the superficial layer at the free surface of the film, which has a particularly complex dynamics \cite{RothFilm,ForrestFilm,sim,JonesFilm,McKennaFilm,NapReview}.
The result is noticeable and suggests that, aside from the adherent layer, the relaxation and the vibrational dynamics of the other layers of the thin film correlate as in the bulk.

\begin{figure}[t]
\centering
\includegraphics[width= 0.9\linewidth]{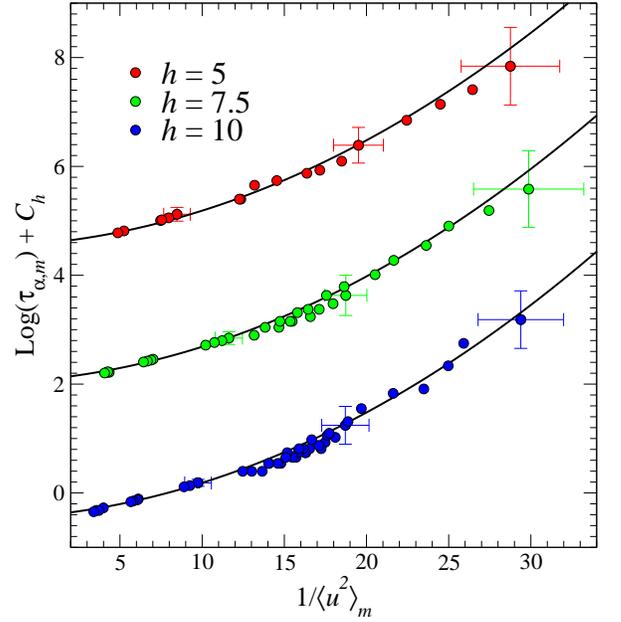}
\caption{Correlation plot of  structural relaxation time and  DW factor of the layers of the film with the indicated thickness at different temperatures ($T=0.47$, $0.48$, $0.49$ and $0.50$). The adherent layer is not included. For a given thickness, the points corresponding to different temperatures and layers have the same colour. For clarity reasons only typical error bars are indicated and the value of $\log \tau_{\alpha,m}$ has been shifted vertically  of a quantity $C_h$ depending on the thickness ($C_5=5$, $C_{7.5}=2.5$ and $C_{10}=0$). The continuous black lines are  Eq.\ref{parabolaMDfilm}  shifted by the same amount.}
\label{fig5}
\end{figure}

\begin{figure}[t]
\centering
\includegraphics[width= 0.9\linewidth]{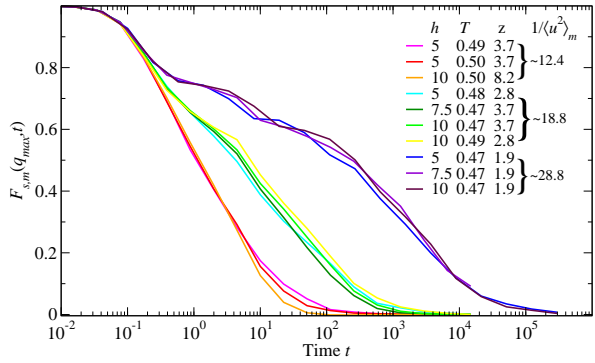}
\caption{Illustration of the predictability of the scaling between vibrational dynamics and relaxation in a thin supported film: the shape of the relaxation function for times not exceeding $\tau_{\alpha, m}$ depends only on the DW factor irrespective of the film thickness, temperature and layer position. The small deviations between the decays observed for times longer than $\tau_{\alpha, m}$ for the states with faster relaxation are due to interlayer mixing.}
\label{fig6}
\end{figure}

To provide a sound basis to the previous result we have investigated films with different thickness and temperature. The results are summarized in Fig. \ref{fig5} for all the layers but the adherent layer. We stress that the MD results are compared to Eq. \ref{parabolaMDfilm} with {\it no adjustement}. We see that structural relaxation and vibrational dynamics of the layers exhibit the same scaling of the bulk system.

The results of Fig. \ref{fig5} suggest that, apart from the adherent layer, layers with equal DW factor $\langle u^2\rangle_m$ exhibit equal relaxation time $\tau_{\alpha, m}$, as stated by Eq.\ref{parabolaMDfilm}. Having defined GT as occurring at a temperature where the relaxation time has a conventional well-defined value, a sharp relation between the DW  factors of the m-th layer of the film evaluated at the corresponding GT  temperature $T_{g, m}$ and the corresponding quantities of the bulk system,  $\langle u_g^2\rangle$ and $T_g$, is predicted  
\begin{equation}
\langle u^2\rangle_m( T_{g, m} ) = \langle u^2\rangle( T_g ) = \langle u_g^2\rangle
\label{filmVSbulk}
\end{equation}
More generally, the results of Fig. \ref{fig5} and the relation between the pair of parameters $(\beta, \gamma)$ and 
$(\tilde{\beta}, \tilde{\gamma})$ \cite{OurNatPhys}, suggest that Eq.\ref{parabolaMDfilm} may be recast in the universal form 
\begin{equation}
\log \tau_{\alpha, m} = \alpha+\tilde{\beta} \, \frac{\langle u_g^2\rangle}{\langle u^2\rangle_m} +\tilde{\gamma}\left (\frac{\langle u_g^2\rangle}{\langle u^2\rangle_m} \right )^2
\label{eqn:u2tauExp3}
\end{equation}
Eq.\ref{eqn:u2tauExp3} is the analogous of Eq. \ref{eqn:u2tauExp2} for thin films. It allows to monitor the solidification of the m-th layer of the film by using solely information concerning  the vibrational properties of the layer and the bulk.

For bulk systems it was shown that particle ensembles with equal DW factor have identical ISF up to the structural relaxation time \cite{OurNatPhys,lepoJCP09,Puosi11}. This a stronger conclusion than the mere scaling between DW and the relaxation time. In an attempt to see if that conclusion may be duly extended to thin films, we wondered if  layers with equal DW factor have identical ISFm up to $\tau_{\alpha, m}$, irrespective of the film thickness, temperature, and layer position.  Fig.\ref{fig6} provides a positive answer in three different mobility regimes.

{  Only the adherent layer fails to  comply with the scaling between vibrational dynamics and structural relaxation. This is  explained by the fact that, broadly speaking, the scaling correlates the local stiffness of the cage, as expressed by the inverse of DW, with the escape rate of the particle trapped in it \cite{OurNatPhys}.  
However, very  close to the substrate, the local stiffness is hardened by the absorption process which superimpose to the one due to the cage effect, resulting in a weaker correlation. The effect is negligible farther from the substrate. 
The presence of an adherent layer is reported by both experiments  \cite{GranickLangm96,NapolitNatCom11} and simulations \cite{BinderPaulThreeStepDecayEPL12}. It is a limited part of the so- called "bound" layer having much lower mobility than the film interior and observed near an attractive substrate in a region with thickness $h_{sub}$ \cite{ZhangDouglasStarrThinSupportedFilmJCP17,ZhangDouglasStarrThinSupportedFilmPNAS18}. Referring to Fig.\ref{fig4} (left), we may estimate $h_{sub}$ as the position of the inflection point limiting the slowed-down region close to the substrate. This yields $h_{sub} \sim 4.1$. This estimate is very close to $h_{sub} \sim 3.7$ derived in Ref.\cite{ZhangDouglasStarrThinSupportedFilmPNAS18} for the same thickness by using a very similar model to ours.} 

\begin{figure}[t]
\centering
\includegraphics[width=0.9 \linewidth]{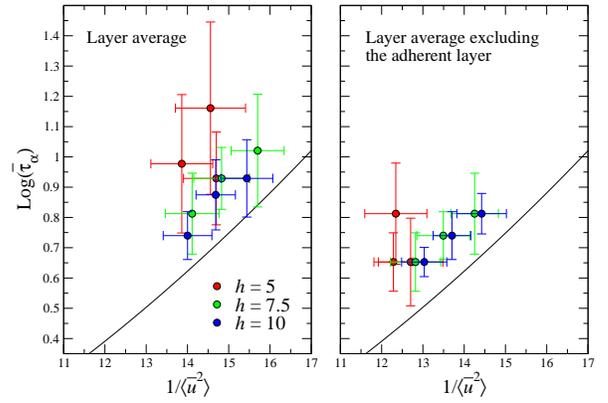}
\caption{{ Missed scaling of the relaxation time and DW factor if averaged over both all the layers (left panel) and all the layers but the adherent one (right panel) of a film. Thicknesses and color code as in Fig.\ref{fig5}.  The black line is Eq.\ref{parabolaMDfilm}. Note that the deviations  increase by decreasing the thickness if the average include {\it all} the layers.}
}
\label{fig7} 
\end{figure}

{ 
Previous studies considered the average fast and vibrational dynamics of the {\it whole} film \cite{DouglasStarrPNAS2015,SimmonsConfined_u2_ACSMacroLett14}. We show that the  scaling tends to be hidden if the dynamics is averaged over the whole film, including or not the adherent layer. To this aim, we first average  the self part of the intermediate scattering function over the particles of interest and consider the time $ \overline{\tau} _{\alpha}$ when the resulting curve drops at 1/e. Then, we perform the average over the same particles to draw the average DW factor $\overline{\langle u^2\rangle}$.   Fig.\ref{fig7} shows that the scaling of the relaxation time and DW factor is missing if averaged over both all the layers (left panel) and all the layers but the adherent one (right panel) of a film. Note that there are  appreciable deviations from  Eq.\ref{parabolaMDfilm} which increase by decreasing the thickness if the average includes {\it all} the layers. Anyway, the deviations, even if appreciable, are not large. This offers an explanation of why the scaling is recovered by small adjustment of a single parameter \cite{DouglasStarrPNAS2015}.}

\section{Conclusions}
We  studied by MD simulations a  class of supported thin films with attractive substrate interaction and different thickness and temperature. The films are analysed by partitioning them into layers as thin as one particle size with the purpose of investigating the observed anisotropic mobility and the strong gradients of both the fast and the slow dynamics. We define a characteristic structural relaxation time of the layer and prove that, aside from the  single layer adherent to the substrate, it exhibits strong correlation with the fast vibrational dynamics of the layer, as accounted for by the DW factor of the particles. We find that the correlation is the same of bulk in the sense that it is described by the same master curve with no adjustable parameters. { Our results suggest that the solidification process of each layer may be tracked  by knowing solely the vibrational properties of the layer and the bulk.}
The scaling is hidden if the average dynamics of the film is inspected.

\section*{Acknowledgements}
Francesco Puosi and Antonio Tripodo are thanked for helpful discussions. We acknowledge the support from the project PRA-2018-34 ("ANISE") from the University of Pisa. A generous grant of computing time from IT Center, University of Pisa and Dell EMC${}^\circledR$  Italia is also gratefully acknowledged.

\bibliography{biblio} 
\bibliographystyle{rsc}

\end{document}